%% file: main.tex
\documentclass[11pt,letterpaper]{article}

\usepackage[utf8]{inputenc}
\usepackage[T1]{fontenc}
\usepackage[margin=1in]{geometry}
\usepackage{graphicx}
\usepackage{booktabs}
\usepackage{tikz}
\usetikzlibrary{positioning}
\usepackage{natbib}
\usepackage[hyphens]{url}
\usepackage{hyperref}
\definecolor{linknavy}{rgb}{0.10,0.10,0.44}
\definecolor{citegreen}{rgb}{0.00,0.30,0.13}
\definecolor{urlblue}{rgb}{0.06,0.20,0.48}
\hypersetup{
  colorlinks=true,
  linkcolor=linknavy,
  citecolor=citegreen,
  urlcolor=urlblue,
  pdfborder={0 0 0}
}
\usepackage{etoolbox}

\apptocmd{\thebibliography}{\raggedright}{}{}

\title{Adoption Telemetry: Measuring Enterprise AI Adoption from Production Signals}
\author{Damon A. Young\\ PolyWise Partners\\ \texttt{dayoung@alumni.princeton.edu}}
\date{\today}

\begin{document}
\maketitle

\begin{abstract}
\input{sections-tex/00-abstract}
\end{abstract}

\input{sections-tex/01-introduction}
\input{sections-tex/02-related-work}
\input{sections-tex/03-adoption-telemetry}
\input{sections-tex/04-nante}
\input{sections-tex/05-implementation}
\input{sections-tex/06-limitations}
\input{sections-tex/07-incumbents}
\input{sections-tex/08-agenda}
\input{sections-tex/09-conclusion}

\bibliographystyle{plainnat}
\bibliography{references}

\end{document}

%% file: sections-tex/00-abstract.tex
We introduce \textbf{adoption telemetry}: a method for measuring enterprise AI adoption by computing change-management stage-progression directly from production usage signals. We contribute (1) a framework unifying pre-deployment evaluation gates, production telemetry, and change-management staging into one instrumented system; (2) \textbf{NANTE}, a concrete five-stage operationalization with defined telemetry thresholds, published openly so they can be tested and disproven; and (3) an open-source reference implementation that distinguishes a healthy cohort from five characteristic adoption-failure modes on synthetic populations with known ground truth. We are explicit that the thresholds are proposed constructs requiring empirical validation against real outcomes --- a research agenda we outline --- not a calibrated model.

%% file: sections-tex/01-introduction.tex
\section{Introduction}\label{introduction}

Enterprise adoption of generative AI has produced an unusual pattern: near-universal deployment alongside near-universal disappointment. The evidence spans three distinct scopes, and we separate them deliberately. On generative-AI pilots broadly, MIT's NANDA initiative --- drawing on executive interviews, employee surveys, and analysis of 300 public deployments --- found that 95\% delivered no measurable P\&L impact \citep{mitnanda2025}.\footnote{A preliminary (v0.1) report whose methodology counts vary across secondary accounts; we cite its headline finding as directional evidence of the failure consensus, not as a precise estimate.} On enterprise AI initiatives at portfolio level, S\&P Global Market Intelligence found the share of companies abandoning most of their initiatives rose from 17\% to 42\% in a single year \citep{spglobal2025vote}. And specifically on \emph{agentic} systems --- those that go beyond augmenting workflows to automating them, operating with autonomy under defined guardrails --- Gartner predicts over 40\% of projects will be cancelled by the end of 2027, citing escalating costs, unclear business value, and inadequate risk controls \citep{gartner2025agentic}.\footnote{A Gartner analyst prediction. The release discloses a January 2025 poll (n{=}3{,}412 webinar attendees) on investment posture, but no derivation for the 40\% cancellation forecast itself; we cite it as directional evidence of the failure consensus specific to agentic systems, not as a measured rate.} These are different populations measured by different methods, and treating them as one statistic obscures more than it shows. What they share is a locus: in each, the reported cause of failure is organizational rather than technical.

This diagnosis is no longer contested. Across the analyst literature, the consulting research, and the trade press, the cause of enterprise AI failure is attributed to organizational factors --- workflow integration, sponsorship, skills, and behavior change --- rather than to model capability. A substantial vocabulary has grown up around the phenomenon: pilot purgatory, pilot fatigue, the GenAI divide, AI theater. We take the diagnosis as established and do not restate it as a finding.

What has not followed the diagnosis is an instrument. Organizations attempting to manage AI adoption today have access to three kinds of measurement, none of which measures adoption in the sense the failure literature describes. Agent evaluation and observability platforms instrument the \emph{system} --- traces, latency, cost, output quality --- and report nothing about the humans invoking it. Enterprise copilot dashboards instrument \emph{activity} --- active users, license utilization, retention rates --- and report how many people touched the tool --- carrying, at most, habit-formation tiers over the same activity counts rather than a model of organizational change or of where it fails. Product analytics contributes a mature tradition of behavioral thresholds, developed to optimize voluntary consumer products rather than to manage a fixed population through an organizational change. Meanwhile the discipline that owns adoption as its subject matter --- change management --- measures it by survey and practitioner assessment: instruments that touch the behavioral record only as coarse usage totals, never as the substrate from which adoption states are inferred.

The result is a measurement gap with a specific shape. An employee who samples an AI assistant twice a week and an employee whose work it has restructured are indistinguishable on a usage dashboard: both are active users. Recent behavioral telemetry suggests this distinction is where the failure lives. Analysis of enterprise AI usage across a large monitored population found that 57\% of AI users spend under 1\% of their working hours in AI tools \citep{activtrak2026sotw}, and quarterly tracking of 120,000+ workers found that while 82\% of AI users sustain usage quarter over quarter, only about 2\% reach a level of use that is consistently embedded in their workflows \citep{activtrak2026maturity}. Breadth is not the problem. Depth is. And depth is precisely what the available instruments cannot interpret.

We propose that this gap be closed by treating adoption as an engineering problem, and we introduce \textbf{adoption telemetry}: the continuous measurement of behavior change from production system signals, interpreted through an explicit model of change. Where observability instruments the system and usage analytics counts activity, adoption telemetry instruments the change itself --- mapping logged behavior to change-management milestones expressed as computable thresholds, so that an organization can locate \emph{where} in the adoption process a population is failing, and act on that location, rather than observe that usage is low and infer a cause.

This paper makes three contributions. First, we describe a framework that unifies three practices currently developed in institutional isolation --- pre-deployment evaluation gates, production usage telemetry, and change-management staging --- into a single instrumented system. Second, we present \textbf{NANTE}, a concrete five-stage operationalization (Notice, Attempt, Navigate, Transform, Embed) that expresses each stage as a defined signature over production telemetry, with published thresholds anyone can test and disprove. Third, we provide an open-source reference implementation and demonstrate that it distinguishes a healthy cohort from five characteristic adoption-failure modes on synthetic populations with known ground truth.

We are explicit about what this evidence does and does not establish. The reference implementation demonstrates \emph{computability}: that change-model stage progression can be derived mechanically from telemetry an enterprise already generates. It does not establish \emph{validity}: the thresholds we propose are unvalidated constructs, not values calibrated against real adoption outcomes. Establishing that calibration requires production data from real deployments, which we outline as a research agenda in Section 8 and for which we invite collaboration.

%% file: sections-tex/02-related-work.tex
\section{Related Work}\label{related-work}

Measurement relevant to AI adoption exists today in four distinct traditions, developed by four professional communities: agent evaluation, enterprise usage analytics, product analytics, and change management. Each instruments an object adjacent to adoption; none instruments adoption itself, in the sense the failure literature uses the term --- the progression of a working population from access to changed behavior. We review each tradition, characterize precisely what it measures, and then examine the structure of the gap between them.

\subsection{Agent evaluation and observability.}\label{agent-evaluation-and-observability.}

The most active measurement tradition in the agent ecosystem evaluates the \emph{system}. AgentBench \citep{liu2023agentbench} assesses agents across interactive environments; \(\tau\)-bench \citep{yao2024taubench} and its successor \(\tau^2\)-bench \citep{barres2025tau2} add task completion under interaction with simulated users, including reliability across repeated trials; a recent survey \citep{yehudai2025survey} organizes the field's evaluation methodologies and benchmarks. Commercial observability platforms extend this instrumentation into production, capturing traces, token costs, latencies, rubric- or model-graded output quality, and, increasingly, per-user usage and end-user feedback signals. This tradition supplies the quality gates a deployment decision requires, and our framework incorporates such gates directly. But its object remains the agent: the human appears as a simulated counterpart, a rater, or a telemetry source --- not as a population progressing through an organizational change. An agent can pass every benchmark and be abandoned by its intended users; the benchmark tradition cannot register that event at all, and production observability registers it only as unexplained telemetry --- falling usage --- disconnected from the quality constructs the same platforms measure and carrying no account of what kind of failure the decline represents.

\subsection{Enterprise usage dashboards.}\label{enterprise-usage-dashboards.}

The second tradition instruments \emph{activity}. Microsoft's Copilot Dashboard, the most widely deployed example, organizes its metrics into four categories --- readiness, adoption, impact, and sentiment --- and defines its central construct precisely: an active Copilot user is one who has performed at least one intentional action in a supported application within the preceding 28-day window \citep{microsoft-copilot-docs}. The dashboard reports adoption trends, per-application usage, total actions, and modeled impact estimates such as assisted hours; a benchmarking layer added in late 2025 compares adoption metrics against internal cohorts and similar organizations; exportable user-level metrics at day and week granularity include per-application intensity measures such as weeks-active-in-28-days. Salesforce provides analogous analytics for Agentforce, and cross-vendor aggregators have begun consolidating usage measurement across AI tools. The raw material in this tradition is genuinely rich --- the exports contain much of the behavioral signal our framework consumes --- and the tradition's strongest artifact goes further. Microsoft's published Copilot Analytics methodology segments users into five mutually exclusive tiers over a rolling twelve-week window --- Non-user, Low, Novice, Habitual, and Power --- where a habitual user is one who used Copilot in at least nine of the past twelve weeks, and a power user is a habitual user averaging fifteen or more weekly Copilot actions \citep{microsoft-segments}.\footnote{The full five-tier segmentation ships in Microsoft's analyst sample-code library; a power-user habit view has additionally been surfacing natively in the Copilot Dashboard during 2026. Microsoft describes the rule as grounded in habit-formation research and its own usage data, without published citation.} This is genuine behavioral staging, and evidence the tradition is reaching toward depth. What separates it from what we propose is the model behind the tiers: they are habit-formation gradations over the same activity counts --- how often, how consistently --- rather than milestones of an organizational change; the segmentation locates a user on a usage gradient, not a population in a change process; and no tier maps to a diagnosis or an intervention. Its cohort comparisons can likewise locate an underperforming population but not characterize what \emph{kind} of stall it represents, and its adoption guidance remains playbook-level rather than a defined mapping from diagnosis to intervention. The dashboards can say more than that usage is low --- they cannot say what kind of problem low usage is.

The closest neighbor to our work sits at the boundary of this tradition. ActivTrak's workforce analytics reports behavioral findings across large monitored populations --- including the finding, central to our motivation, that 82\% of AI users sustain usage quarter over quarter while only 2\% reach a stage where AI is consistently embedded in workflows \citep{activtrak2026maturity} --- and has recently introduced a three-stage behavioral maturity classification. This is behavioral staging computed from telemetry, and we regard it as convergent evidence that the depth of use, not its breadth, is the discriminating variable. It differs from what we propose in three respects: its stages are descriptive usage tiers rather than milestones of an explicit change model; it publishes no explicit stage-to-intervention mapping; and it is bound to a proprietary monitoring platform rather than offered as an open, auditable instrument over an organization's own agent telemetry.

\subsection{Product analytics.}\label{product-analytics.}

The third tradition owns the deepest expertise in behavioral thresholds --- applied from a different seat. Growth practice identifies \emph{activation} events --- the earliest behaviors separating retained users from churned ones --- historically through retrospective correlation and judgment, increasingly through experimental validation: Slack's canonical threshold held that a team which had exchanged 2,000 messages had ``really tried'' the product, after which, by its founder's account, 93\% of such teams were still active \citep{firstround-slack}. Cohort retention curves, activation funnels, and north-star metrics constitute a mature instrumentation of behavior over time, and our framework borrows these methods directly. In its self-serve and product-led core, this tradition optimizes the engagement of users who opted in, in service of the vendor's retention and revenue. Its adjacent practices complicate the seat, though not the conclusion. Customer-success instrumentation --- seat activation, license utilization, health scores --- does track fixed, employed populations, but from the vendor's side of the table, per account, in service of the renewal. The digital-adoption-platform category sits differently: platforms such as WalkMe --- acquired by SAP for approximately \$1.5 billion in 2024 \citep{sap-walkme} --- are sold to the adopting enterprise itself, instrument the organization's whole software estate, and now market AI-adoption measurement in terms this paper would endorse: ``You know how many licenses you bought. You don't know if anyone is using them for real work'' \citep{walkme-home}. What the category supplies is instrumentation without a model: in-application guidance and friction analytics oriented to task completion within tools, with no computed staging of a population through a change model, no taxonomy of stall types, no mapped interventions, and no open, disputable thresholds. What the tradition carries little account of is the \emph{organizational} layer --- sponsorship, incentives, workflow authority --- that determines whether an employed population's work actually changes, measured from the adopting organization's side of the table rather than the vendor's.

\subsection{Change management.}\label{change-management.}

The fourth tradition owns the theory our framework operationalizes. Its academic lineage runs from the Technology Acceptance Model --- perceived usefulness and perceived ease of use \citep{davis1989} --- through UTAUT's consolidation of the acceptance determinants into performance expectancy, effort expectancy, social influence, and facilitating conditions \citep{venkatesh2003}, to recent extensions addressing generative AI in the workplace \citep{wolfe2025}. These models identify the determinants of individual acceptance with validated questionnaire instruments; where the behavioral record appears --- as in UTAUT's own logged-usage criterion variable --- it serves as a coarse usage outcome, not as a substrate from which adoption states are inferred. The field's own critique anticipated ours: the post-adoptive literature argued two decades ago that adoption research stops at the moment of first use, and called for feature-level accounts of what individuals do afterward \citep{jasperson2005} and for reconceptualizing ``system usage'' itself beyond lean frequency counts \citep{burtonjones2006} --- a program whose instruments, however, remained surveys. Staged accounts of technology assimilation exist in the implementation literature as well \citep{cooper1990}; what the acceptance models lack, practitioner frameworks supply in operational form --- most prominently the ADKAR model, whose five sequential elements (Awareness, Desire, Knowledge, Ability, Reinforcement) give change a staged, milestone structure with the premise that different stall points demand different interventions \citep{hiatt2006}. NANTE shares that premise rather than ADKAR's vocabulary: its stages --- Notice, Attempt, Navigate, Transform, Embed --- deliberately echo none of Awareness, Desire, Knowledge, Ability, Reinforcement. This machinery has no equivalent in the other three traditions. Its instrumentation, however, measures by asking: surveys and structured, expert-mediated assessments --- episodic and sample-limited. Behavioral data is not absent from the practice --- practitioner guidance tracks utilization and proficiency through usage reports and system data, and change teams routinely consult adoption dashboards --- but it enters as corroborating totals rather than as measurement of position in the model. The discipline's newest instruments hold to the method: Prosci's AI Adoption Diagnostic is a structured, expert-mediated assessment across five diagnostic dimensions, grounded in survey research with over 1,100 professionals \citep{prosci-diagnostic}. Perception measures capture constructs telemetry cannot, and we position them as complements. The gap is precise: the discipline that knows what adoption \emph{is} has no validated method for inferring its own stage milestones from the continuous behavioral record --- no existing practice computes, from what a population actually does, where in the staged model that population stands.

\subsection{The structure of the gap.}\label{the-structure-of-the-gap.}

These four traditions are developed by four communities --- machine-learning and HCI researchers, platform product organizations, growth practitioners, and organizational-change professionals --- that publish in different venues, sell to different buyers, and, as far as we can determine, rarely cite one another's methods and never adopt one another's constructs. One methodological ancestor deserves separate note. Process mining infers how work is actually performed --- and how it conforms to an intended model --- from event logs \citep{vanderaalst2016}, and is in that sense the nearest methodological relative of what we propose: the founding move, treating the event log as the authoritative record of organizational behavior, is the same. The object differs --- process conformance rather than a population's progression through a behavioral change --- and so does the interpretive layer: process models rather than change models. Elements of the human signal do cross these boundaries --- end-user feedback flows into observability platforms, behavioral staging appears in workforce analytics, employed populations are instrumented by adoption platforms --- but no tradition unifies deployment-quality gates with the stage-progression of a managed population's behavior change under one measurement framework, which is the specific unification this paper proposes. The gap between the traditions is therefore structural rather than incidental, a point we return to in Section 7. Its consequence is a specific blindness: no open instrument computes the stage-progression of a managed population's behavior change from production telemetry through an explicit model of change with a mapped intervention response. The elements exist in fragments across the four traditions; the unification does not. The trajectory of the field has made this blindness newly consequential. Toxtli's 2024 position paper on Human-Centered Automation argued that automation fails its users through expertise demands and opaque interfaces, and prescribed natural-language, accessible, user-centered automation \citep{toxtli2024hca}. That prescription substantially won: conversational interfaces became the dominant automation paradigm, and building agents ceased to require specialist expertise. Its victory relocated the constraint. When automation was hard to build and hard to use, usability was the binding limit on adoption; now that agents are easy to build and easy to use, the binding limit is whether work actually changes --- a variable that none of the four traditions above was constructed to observe. Instrumenting it is the subject of the remainder of this paper.

%% file: sections-tex/03-adoption-telemetry.tex
\section{Adoption Telemetry}\label{adoption-telemetry}

We define \textbf{adoption telemetry} as the continuous measurement of behavior change from production system signals, interpreted through an explicit model of change. The definition carries three commitments. The \emph{object} of measurement is a working population's progression through an organizational change --- not the system's quality, and not raw activity. The \emph{substrate} is telemetry the deployed systems already emit --- invocations, sessions, turns, task outcomes --- requiring no surveys, no interruption of work, and no additional instrumentation of people. The \emph{interpretation} is an explicit, staged model of change whose milestones are expressed as computable predicates over that telemetry, so that every classification the instrument produces is auditable down to the signals that triggered it and falsifiable by anyone who disputes the thresholds.

Each commitment marks a contrast with one of the four traditions of Section 2. Against agent evaluation and observability, adoption telemetry changes the \emph{object}: it instruments the population, not the agent, and treats a passing benchmark as a precondition for measurement rather than its conclusion. Against enterprise usage dashboards, it changes the \emph{construct}: the same activity counts appear, but as evidence for position in a model of change rather than as the measurement itself --- the difference between reporting that usage is low and identifying which stage of adoption has failed. Against product analytics, it changes the \emph{seat}: the instrument serves the adopting organization's change objective rather than a vendor's renewal objective, measuring one organization's whole population across its tools rather than one vendor's product across its accounts. Against change-management practice, it changes the \emph{method}: adoption states are inferred from what the population does rather than assessed from what it reports, converting the discipline's staged models from survey constructs into computed ones.

An instrument qualifies as adoption telemetry, in our usage, when it satisfies four properties. It is \emph{computable}: every stage classification derives mechanically from logged behavior, with no expert judgment in the loop at measurement time. It is \emph{model-explicit}: the stages, their ordering, and the thresholds that operationalize them are stated openly enough to be criticized, not embedded implicitly in a scoring function. It is \emph{population-level}: classification applies to cohorts, and the instrument's outputs are distributions and stall points, not individual dossiers. And it is \emph{intervention-mapped}: each diagnosable stall corresponds to a stated class of response, so that measurement terminates in action rather than in reporting.

The population-level property is a design constraint, not a convenience. Adoption telemetry as we define it measures cohorts --- teams, functions, rollout waves --- and never individuals. This is partly principled: the object of interest is organizational change, which is a property of populations, and per-person adoption scoring would convert a change instrument into a surveillance instrument, corroding the trust on which honest usage data depends. It is also practical: cohort-level aggregation with minimum-size floors is what makes the instrument deployable in enterprises whose works councils, privacy offices, and security teams would correctly refuse individual behavioral scoring. The instrument's unit of diagnosis is where a \emph{population} stalls, not who is failing. The substrate narrows the exposure further: adoption telemetry consumes records of interaction with the deployed system itself --- computed behavior rather than self-report, and not ambient workplace monitoring --- so the instrument observes only what the population does with the capability being measured.

\textbf{Scope: what class of system.} Adoption telemetry as specified here assumes that a human initiates the interactions being counted, and that assumption holds unevenly across the systems now called AI. It holds cleanly for \emph{assistants} --- copilots and chat interfaces where every session begins with a human prompt. It holds, with one extension we specify in Section 8.5, for \emph{human-in-the-loop agents} --- systems that execute multi-step work under human initiation and review, where the interesting question shifts from whether the population invokes the agent to whether it lets the agent act. It does not hold for \emph{autonomously operating agents} --- systems that, once configured, execute workflows on schedules or triggers without per-task human initiation. For that third class the framework's central signal inverts: deepening adoption \emph{reduces} human-initiated invocation, so an organization succeeding at delegation and one abandoning the tool produce the same declining curve, and NANTE as specified would diagnose the first as reinforcement decay and prescribe re-engagement. We regard this as a genuine boundary of the present work rather than a defect to be patched: the unit of analysis for a fully delegated workflow is plausibly not a population of users progressing through a change at all, but an organization transferring ownership of a process --- a different construct, requiring a different instrument.

Architecturally, adoption telemetry unifies three layers that Section 2 showed are currently built and operated in isolation. A \emph{readiness layer} carries the evaluation gates of the agent-quality tradition: an agent enters measurement only after passing the capability and safety bars appropriate to its task, because adoption measurement of a failing agent measures nothing but the population's good judgment in avoiding it. A \emph{signal layer} carries the behavioral record: the invocation, session, and outcome events that production systems and enterprise dashboards already emit, ingested in whatever export form the deployment provides. An \emph{interpretation layer} --- the layer no existing tradition supplies --- maps the signal layer to the milestones of an explicit change model, classifies each cohort's position, detects characteristic failure signatures, and emits the corresponding intervention class. The framework's claim is not that any layer is individually novel; the readiness and signal layers exist today as separate industries. The claim is that the interpretation layer, and the unification of the three, converts data that organizations already possess into a diagnosis they currently cannot make.

What this makes possible is a change in the question an organization can ask. Usage dashboards answer \emph{how much is it being used}; evaluation platforms answer \emph{does it work}; surveys answer \emph{how do people feel about it}. Adoption telemetry answers \emph{where, in the process of changing how work is done, has this population stopped --- and what kind of problem is that}. Section 4 presents NANTE, our operationalization of this definition.

%% file: sections-tex/04-nante.tex
\section{NANTE}\label{nante}

NANTE operationalizes adoption telemetry as a five-stage model --- Notice, Attempt, Navigate, Transform, Embed --- with each stage defined as a computable predicate over the event stream a deployed agent already produces.\footnote{The name is drawn from the Twi \emph{nante} --- ``walk'' --- as in \emph{nante yiye}, ``walk well'': the model measures a population's walk through a change, stage by stage.} We present the stages, the telemetry signatures that classify them, the detection of stalls and characteristic failure signatures, and the mapping from diagnosis to intervention. Throughout, the specific threshold values are \emph{proposed defaults}: illustrative starting points chosen to produce statistically distinguishable classifications on synthetic populations, published openly in the reference implementation precisely so that they can be disputed, tuned, and eventually calibrated against real outcomes. No value below is a validated finding.

\subsection{The staged model.}\label{the-staged-model.}

Each stage answers one question about a user's relationship to the deployed capability. \emph{Notice}: does the user know the capability exists and applies to them? \emph{Attempt}: have they tried it at all? \emph{Navigate}: has trial become recurring use? \emph{Transform}: has recurring use become integrated, multi-step work with acceptable success --- has the shape of their work changed? \emph{Embed}: is that integration continuous enough that withdrawal would disrupt output? The first three stages concern breadth --- whether people arrive at the tool. The last two concern depth --- whether work changes. The published evidence of Section 1 locates the characteristic enterprise failure at precisely the breadth-to-depth boundary, and the model is built to make that boundary observable.

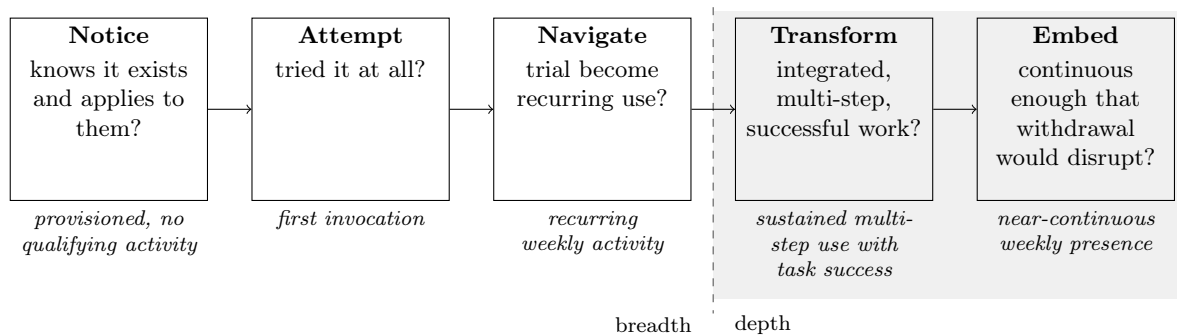
\begin{figure}[htbp]
  \centering
  \input{figures/nante-model.tikz}
  \caption{The five-stage NANTE model: each stage, the question it answers, and its telemetry signature. Embed's continuity requirement applies within the population already clearing Transform's gates (\S4.2).}
  \label{fig:nante-model}
\end{figure}

\subsection{Telemetry signatures.}\label{telemetry-signatures.}

The substrate is an event stream of invocations with session identifiers, per-session turn counts, and task outcomes, joined to a \emph{provisioned-user roster}. The roster is a structural requirement, not a convenience: a store containing only events renders Notice invisible, because the users who never arrived generate no rows --- the population that most needs counting is exactly the one that leaves no trace. Enterprise admin tooling does count provisioned-but-inactive users in readiness reports; the roster requirement makes that same population native to the measurement model itself, where its members receive a stage, a diagnosis, and a mapped response rather than appearing as a license statistic. With the roster in place, the signatures are: a user classifies as \textbf{Notice} if provisioned but without qualifying activity; \textbf{Attempt} on first invocation; \textbf{Navigate} at recurring use, proposed as activity in at least 3 distinct weeks; \textbf{Transform} at sustained multi-workflow integration, proposed as at least 6 active weeks, a 60\% share of sessions that are multi-step, and a task success rate of at least 55\%; \textbf{Embed} at sustained continuity, proposed as activity in at least 90\% of the weeks since the user's first use, so that historical volume cannot substitute for present integration. One consequence of these definitions deserves explicit statement: Transform and Embed partition the population that clears Transform's gates --- by continuity --- so an Embed classification implies Transform's criteria. The stages above Navigate are not successive filters but a split of one qualifying pool, and per-user classification assigns the deepest stage whose predicate is satisfied. Two proxies deserve note. ``Multi-step'' is proposed as sessions of 12 or more turns --- an honest proxy for workflow depth in an event schema that carries no task-type field. And the Transform success-rate gate is a deliberate addition beyond a literal reading of ``sustained use across several workflows'': without it, a cohort of frequent, multi-step, but mostly \emph{failing} users would classify as transformed, which no defensible construct of adoption should permit.

\subsection{Stall detection and honest ceilings.}\label{stall-detection-and-honest-ceilings.}

A cohort's stall point is located by walking the stage boundaries in order and identifying the first genuine failure of graduation --- the first boundary where the fraction of the roster progressing falls below a proposed floor. For the early boundaries (Notice\(\rightarrow\)Attempt, Attempt\(\rightarrow\)Navigate), graduation of 90\% or more is proposed as healthy and below 80\% as failing. At the two depth boundaries (Navigate\(\rightarrow\)Transform and Transform\(\rightarrow\)Embed) the proposed cutoffs are set apart deliberately: the healthy floor is placed above 1.0 --- a graduation fraction no cohort can reach, so none reads ``healthy'' --- while the at-risk floor is a graduation fraction of 0.01, beneath which a cohort reads ``failing'' for falling below the near-universal \textasciitilde2\% workflow-integration baseline the published evidence reports. The two depth boundaries are treated differently, and the difference is a design decision we make explicitly rather than bury in configuration. The published evidence indicates that progression past Navigate is rare \emph{everywhere} --- the \textasciitilde2\% workflow-integration figure is an industry-wide cliff, not a cohort-specific defect. Classifying an ordinary cohort as ``failing'' at a boundary that nearly every organization fails would make the instrument an alarm that is always ringing. NANTE therefore sets the depth boundaries so that no cohort reads ``healthy'' there --- \emph{at-risk is the honest ceiling} at the breadth-to-depth cliff, reflecting that even good cohorts stand before a hard wall --- and reserves ``failing,'' and therefore the stall-point designation, for cohorts performing meaningfully below the published baseline itself. The instrument's verdicts are thereby calibrated to the field's actual distribution rather than to an aspiration no one meets.

\subsection{Failure signatures.}\label{failure-signatures.}

Beyond the stall point, four characteristic signatures are computed as flags. \emph{Shallow plateau}: breadth-versus-depth divergence --- high and stable participation with flat depth metrics, the index summing a retention component and a session-depth-flatness component (each on the unit interval) and subtracting one, so it is positive only when retention is high \emph{and} session depth stays flat, and negative otherwise --- proposed as a divergence index above 0.3; this is the signature of the cohort every usage dashboard reports as successful. \emph{Champion dependency}: concentration of activity in few hands, proposed as a Gini coefficient above 0.5 over per-user activity among active users --- the never-arrived are excluded, their absence being already counted at Notice; the cohort's apparent adoption is one resignation away from collapse. \emph{Usage regression}: a late-window invocation rate below 40\% of an established early-window rate, gated by preconditions ensuring the cohort had genuinely adopted before declining. We name this flag for what it measures --- regression of usage frequency --- rather than what we would prefer to measure (regression from achieved depth), a distinction the current event schema cannot support and which we carry to the limitations of Section 6. \emph{Low task success}: a Navigate stall driven by the failing success sub-gate rather than by session depth --- the population attempts integrated, multi-step work and fails at it, proposed as at least 75\% of the cohort's Navigate-classified, tenure-eligible users failing the Transform success-rate gate while at most 90\% fail its multi-step-share gate, so that shallow depth does not itself account for the stall --- computed only when the cohort stalls at Navigate and at least 20 such users are evaluated. This is the signature separating a population that never goes deep from one that goes deep and fails, and its name follows the same convention as usage regression: it is named for what is measured.

\subsection{From diagnosis to intervention.}\label{from-diagnosis-to-intervention.}

Each stall point maps to a class of response, and --- because misdirected investment is the modal enterprise failure --- each mapping states a \emph{not-this} contrast naming the intervention the stall does not call for. A \textbf{Notice} stall is an awareness and provisioning failure: the response is visibility and access, \emph{not training} --- one cannot train people to use a tool they do not know exists. One honest limit of this mapping deserves statement: a zero-activity row is behaviorally ambiguous between a user who never learned the capability exists and one who knows and has declined it, and telemetry alone cannot separate the two. The visibility-and-access response is first-line because it is cheap and its premise is verifiable; where a Notice stall persists after access and visibility are confirmed, the population is likely refusing rather than unaware --- a distinction that belongs to the perception instruments of Section 8.6, not to the event stream. An \textbf{Attempt} stall is a motivation failure: the response is role-specific proof of value, \emph{not feature tours} --- the population knows the tool exists and has declined it. A \textbf{Navigate} stall carrying the shallow-plateau signature is a workflow-fit failure: the response is redesigning the workflow so the tool sits inside the actual task sequence, \emph{not more licenses} --- the population already has access; access is not the constraint. A \textbf{low-task-success} signature calls for skill-building and enablement --- worked examples, coaching, task-fit tuning --- \emph{not workflow redesign}: the population has already brought the tool into deep work; the work is failing, not absent. A \textbf{champion-dependency} signature calls for deliberate de-concentration --- pairing the champions with peers, documenting their workflows, and spreading responsibility so adoption does not ride on a handful of people --- \emph{not celebrating the power users} whose concentration is the risk: the cohort's capability must outlive its carriers. Usage-regression signatures call for visible leadership sponsorship and re-integration support, \emph{not re-onboarding} --- the population once knew how; something stopped rewarding the behavior. Where a cohort carries both a stall point and a signature flag with distinct mapped responses --- a population reading as an Attempt stall while flying the usage-regression flag, for instance --- the signature takes precedence: the stall locates the population's current position, but a signature identifies the process that produced it, and treating the process outranks treating the position. The mapping is deliberately coarse --- classes of response, not prescriptions --- because the instrument's job ends where organizational judgment begins: it identifies which kind of problem the cohort has.

\subsection{The composite score.}\label{the-composite-score.}

For executive comparison across cohorts and over time, NANTE reduces the stage distribution to a 0--100 score by ordinal stage weights (proposed: 0, 25, 50, 75, 100), reported alongside --- never instead of --- the distribution, the stall point, and the flags. Table~1 makes the score's insufficiency concrete: the shallow-plateau and ability-gap cohorts share an identical stage distribution and an identical score of 49.9 while carrying opposite diagnoses --- one shallow but succeeding, one deep but failing --- and the awareness-gap and reinforcement-decay cohorts score within 0.6 points (41.0 and 41.6) despite representing a population that never arrived and one that arrived and left. The scalar is order-preserving for executive comparison and provably inadequate for diagnosis, which is exactly why it is never reported without the distribution, stall point, and flags that separate these cases. Where the observation window or cohort size falls below proposed floors (60 post-rollout days and 50 users, with per-metric minimums), the instrument reports the affected values as unavailable rather than computing on insufficient evidence.

%% file: figures/nante-model.tikz
\def\NanteStage#1#2{\parbox[c][2.2cm][t]{2.4cm}{\centering
  \hyphenpenalty=10000\exhyphenpenalty=10000
  {\bfseries #1}\\[2pt]#2\par}}
\begin{tikzpicture}[
  font=\footnotesize,
  stage/.style={draw, line width=0.4pt, inner sep=3pt},
  signal/.style={text width=2.7cm, align=center, font=\scriptsize\itshape,
                 anchor=north, inner sep=1pt},
]
  \def\cA{0} \def\cB{3.2} \def\cC{6.4} \def\cD{9.6} \def\cE{12.8}
  \def\divx{8.0}

  \fill[black!6] (\divx,1.3) rectangle (14.3,-2.5);
  \draw[black!55, dashed, line width=0.5pt] (\divx,1.35) -- (\divx,-2.7);

  \node[stage] (n) at (\cA,0) {\NanteStage{Notice}{knows it exists and applies to them?}};
  \node[stage] (a) at (\cB,0) {\NanteStage{Attempt}{tried it at all?}};
  \node[stage] (v) at (\cC,0) {\NanteStage{Navigate}{trial become recurring use?}};
  \node[stage] (t) at (\cD,0) {\NanteStage{Transform}{integrated, multi-step, successful work?}};
  \node[stage] (e) at (\cE,0) {\NanteStage{Embed}{continuous enough that withdrawal would disrupt?}};

  \draw[->, line width=0.4pt] (n) -- (a);
  \draw[->, line width=0.4pt] (a) -- (v);
  \draw[->, line width=0.4pt] (v) -- (t);
  \draw[->, line width=0.4pt] (t) -- (e);

  \node[signal] at (\cA,-1.3) {provisioned, no qualifying activity};
  \node[signal] at (\cB,-1.3) {first invocation};
  \node[signal] at (\cC,-1.3) {recurring weekly activity};
  \node[signal] at (\cD,-1.3) {sustained multi-step use with task success};
  \node[signal] at (\cE,-1.3) {near-continuous weekly presence};

  \node[anchor=east, font=\scriptsize] at (\divx-0.15,-2.85) {breadth};
  \node[anchor=west, font=\scriptsize] at (\divx+0.15,-2.85) {depth};
\end{tikzpicture}

%% file: sections-tex/05-implementation.tex
\section{Implementation}\label{implementation}

\subsection{The toolkit.}\label{the-toolkit.}

We provide an open-source reference implementation, \emph{agent-adoption-kit}, released under Apache 2.0. It comprises four components: a minimal event schema (invocations with session identifiers, per-session turn counts, and task outcomes) joined to a provisioned-user roster, persisted in SQLite; a synthetic-population simulator that generates event streams for cohorts with configurable behavioral profiles; the analytics engine implementing the NANTE classification of Section 4 --- per-user staging, cohort stall-point detection, failure-signature flags, intervention selection, and the composite score --- with every threshold externalized to a single open configuration file in which each value is marked as proposed; and a report generator that renders a two-cohort comparison as a self-contained HTML page. The implementation covers the framework's signal and interpretation layers; the readiness layer's evaluation gates are specified as preconditions (Section 3) but their harness is not yet part of the toolkit --- a gap we record in Section 8. Three implementation commitments bear on the framework's claims. All classification is deterministic computation over the event stream: a large language model is used only to draft narrative commentary in reports, never in measurement, so every stage assignment and flag is exactly reproducible and auditable to its inputs. All analysis is cohort-level with minimum-size floors, per the design constraint of Section 3. And the repository contains no client data of any kind; every result below derives from synthetic populations generated by code in the same repository, from a fixed random seed.

\subsection{Evaluation method.}\label{evaluation-method.}

The claim under test is computability with discrimination: that NANTE's stage classifications, stall points, and failure signatures can be derived mechanically from the event stream alone, and that the derived diagnoses distinguish the characteristic failure modes the adoption literature describes. We evaluate against synthetic populations with known ground truth. The simulator implements six behavioral profiles --- one healthy and five pathological --- each a generative model of a documented failure pattern: \emph{shallow plateau} (broad, sustained, but persistently shallow use --- the pattern behind the published \textasciitilde2\% workflow-integration figures; intended diagnosis: Navigate stall, shallow-plateau flag); \emph{awareness gap} (a large provisioned-but-never-arriving segment; intended: Notice stall, no flags); \emph{ability gap} (frequent, effortful, but persistently unsuccessful use; intended: Navigate stall, low-task-success flag); \emph{reinforcement decay} (established usage frequency declining across the observation window; intended: Attempt stall, usage-regression flag); and \emph{champion dependency} (activity concentrated in a small fraction of the roster; intended: Attempt stall, champion-dependency flag). These intended diagnoses are the evaluation's answer key, fixed before the engine runs. Each pathology's reference cohort comprises 167 provisioned users observed over 149 days. Evaluation asks whether the engine, given only the emitted events and roster, recovers each profile's intended diagnosis --- the correct stall point and the correct flags --- without access to the generating labels. These known-answer checks run as the repository's test suite in continuous integration.

\subsection{Results.}\label{results.}

\begin{figure}[htbp]
  \centering
  \includegraphics[width=\textwidth]{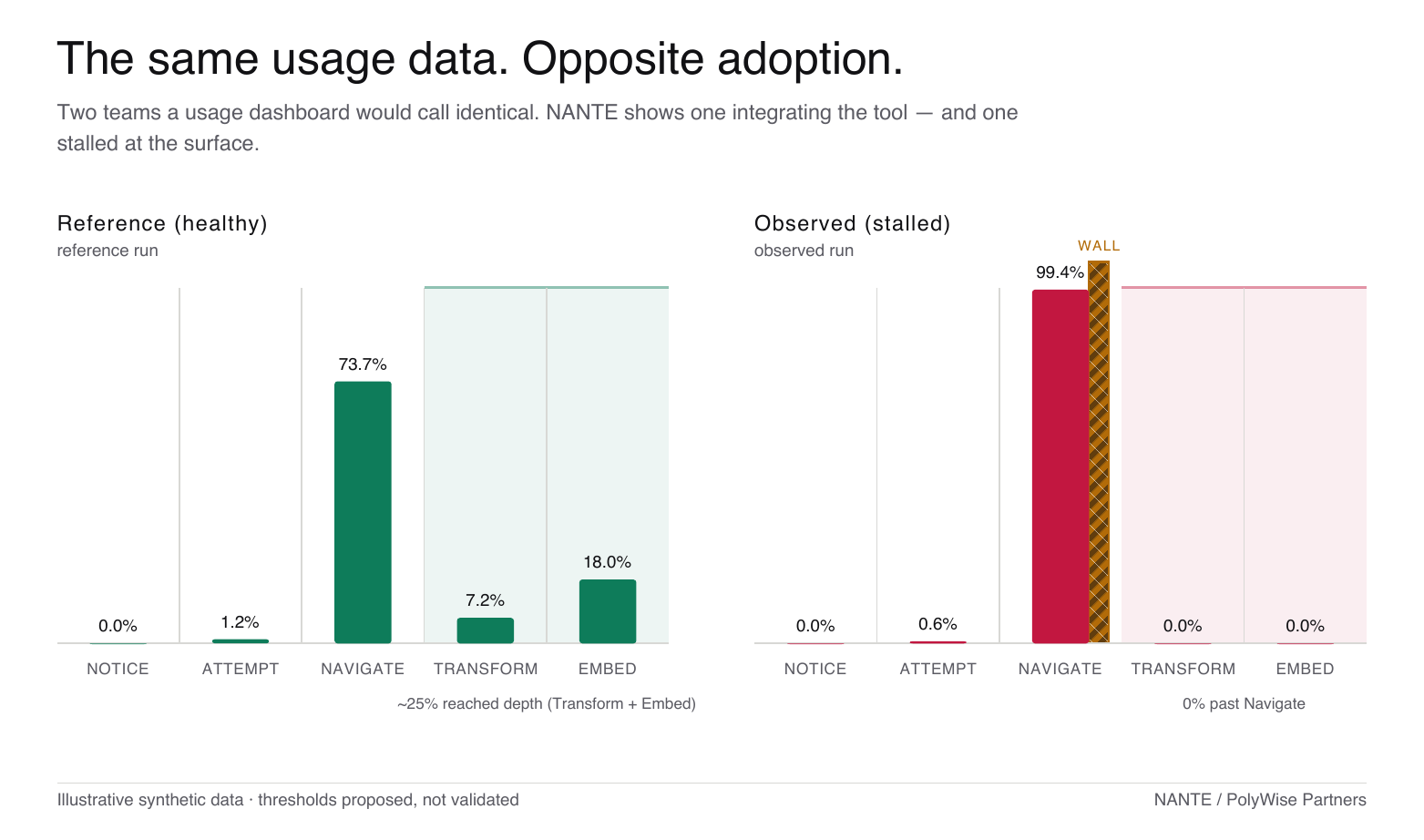}
  \caption{Comparison one-pager for the headline pair, generated by the toolkit (\texttt{-{}-theme=paper} rendering): the same five-stage axis for both cohorts.}
  \label{fig:comparison}
\end{figure}

The engine distinguishes all six profiles. The headline contrast is the pair the framework exists to separate: a healthy cohort and a shallow-plateau cohort that are nearly indistinguishable on breadth --- in the reference run, 98\% or more of both rosters progress into recurring use --- and opposite in depth. The healthy cohort's population distributes upward (73.7\% at Navigate, 25.1\%\footnote{Depth is computed from unrounded stage fractions; it can differ from the sum of the displayed rounded columns by 0.1.} reaching Transform or Embed), scores 60.5, and exhibits no stall; the shallow-plateau cohort concentrates 99.4\% of its population at Navigate with none beyond, scores 49.9, stalls at Navigate, and raises the shallow-plateau flag --- its breadth-versus-depth divergence index (0.67) more than doubles the proposed 0.30 threshold, and the flag's firing is auditable to that number. A usage dashboard reporting active users would rate the second cohort marginally \emph{higher} than the first. The remaining pathologies exercise the framework's distinguishing machinery: the awareness-gap profile stalls at Notice --- a diagnosis possible only because the roster makes never-arriving users countable; the ability-gap profile, whose users are frequent and multi-step but predominantly failing, is barred from Transform by the success-rate gate and reads as a Navigate stall carrying the low-task-success flag: its cohort task success (34.5\%) is half the healthy cohort's (69.5\%), while the shallow-plateau cohort's (69.6\%) is essentially identical to the healthy cohort's --- shallow stalling is not failing, and failing is not shallowness, and the two signatures never co-fire on these populations; the champion-dependency profile raises the champion-dependency flag on the Gini criterion (0.65 against the proposed 0.5 threshold); and the reinforcement-decay profile raises the usage-regression flag on the declining-frequency criterion (0.01 against the proposed 0.4 floor). Full per-pathology results --- stage distributions, scores, stall points, and flags for all six profiles --- are emitted directly by the test suite into the table below, so that the paper's central evidence is mechanically regenerated from the published code rather than transcribed. Figure~\ref{fig:comparison} reproduces the toolkit's generated comparison report for the headline pair: the same five-stage axis for both cohorts, the healthy population visibly filling the Transform-and-Embed band that stands empty behind the stalled cohort's wall at Navigate.

\begin{table}[htbp]
  \centering
  \caption{Per-pathology results emitted by \texttt{make results} from the test suite: stage distribution, cohort size, divergence index, depth, task success rate, composite score, stall point, and flags for all six profiles.}
  \label{tab:pathologies}
  \input{tables/pathologies.tex}
\end{table}

\subsection{What this evaluation establishes --- and what it cannot.}\label{what-this-evaluation-establishes-and-what-it-cannot.}

These results demonstrate that change-model stage progression, stall location, and characteristic failure signatures are computable from telemetry an enterprise deployment already produces, and that the computation discriminates among distinct, literature-grounded failure modes at realistic cohort sizes. They do not demonstrate validity on real populations, and we state the circularity plainly: the synthetic pathologies and the proposed thresholds were developed together, so the evaluation shows that the instrument separates the failure modes it was designed to separate --- a necessary condition and a genuine engineering result, but not evidence that real cohorts classified by these thresholds are correctly diagnosed, nor that the proposed values are calibrated to real-world stage boundaries. The demonstration is also specific to the reference configuration: a single seed and the proposed default thresholds; cross-seed variance and threshold-sensitivity characterization remain part of the agenda of Section 8.3. That evidence can come only from production deployments with observed outcomes, which is the validation agenda of Section 8. Section 6 details these and the framework's other limitations.

%% file: tables/pathologies.tex
\setlength{\tabcolsep}{2.2pt}%
\scriptsize
\begin{tabular}{l rrrrr r r r r r l p{1.5cm}}
\toprule
 & \multicolumn{5}{c}{Stage distribution (\%)} & & & & & & & \\
\cmidrule(lr){2-6}
Pathology & Notice & Attempt & Navigate & Transform & Embed & $N$ & Div. & Depth & Succ. & Score & Stall & Flags \\
\midrule
Healthy & 0.0 & 1.2 & 73.7 & 7.2 & 18.0 & 167 & -0.43 & 25.1 & 69.5 & 60.5 & --- & {\raggedright ---\par} \\
Shallow plateau & 0.0 & 0.6 & 99.4 & 0.0 & 0.0 & 167 & 0.67 & 0.0 & 69.6 & 49.9 & Navigate & {\raggedright Shallow plateau\par} \\
Awareness gap & 32.9 & 1.2 & 49.7 & 1.2 & 15.0 & 167 & -0.59 & 16.2 & 69.7 & 41.0 & Notice & {\raggedright ---\par} \\
Ability gap & 0.0 & 0.6 & 99.4 & 0.0 & 0.0 & 167 & -0.32 & 0.0 & 34.5 & 49.9 & Navigate & {\raggedright Low task success\par} \\
Reinforcement decay & 6.6 & 23.4 & 67.1 & 3.0 & 0.0 & 167 & -0.99 & 3.0 & 69.5 & 41.6 & Attempt & {\raggedright Usage regression\par} \\
Champion dependency & 9.6 & 47.3 & 39.5 & 0.0 & 3.6 & 167 & -0.88 & 3.6 & 69.6 & 35.2 & Attempt & {\raggedright Champion dependency\par} \\
\bottomrule
\end{tabular}

%% file: sections-tex/06-limitations.tex
\section{Limitations}\label{limitations}

We group the framework's limitations into five categories: the validity of its constructs, the fidelity of its evidence, the limits of its measurement schema, the boundaries of its scope, and the behavioral effects of measurement itself.

\subsection{Construct validity.}\label{construct-validity.}

The central limitation has been stated throughout and bears consolidation. Every threshold in NANTE is a proposed default, not a calibrated value: no evidence presented here establishes that three active weeks is where recurring use begins, that a 60\% multi-step share marks workflow integration, or that the stage boundaries as operationalized correspond to the milestones of behavior change they are named for. The evaluation of Section 5 compounds this: as stated there, the pathologies and thresholds were co-developed, and the demonstrated discrimination therefore establishes computability, not correctness. Until stage classifications are validated against independently observed outcomes --- did cohorts classified as embedded in fact sustain changed work practices; did diagnosed stall types respond to their mapped interventions --- NANTE's outputs should be treated as structured hypotheses about a cohort, not findings.

\subsection{Fidelity of evidence.}\label{fidelity-of-evidence.}

All results derive from synthetic populations, and synthetic populations are tidy. Each simulated cohort expresses one pathology; real cohorts plausibly express several at once, at varying intensities, entangled with seasonality, reorganizations, tool changes, and attrition. The simulator draws behavior from parameterized distributions that mirror published aggregate patterns, but no claim is made that its generative processes match the microstructure of real usage. One profile is exempt from this mirroring by design: the healthy cohort's depth conversion (25.1\%) sits roughly an order of magnitude above the $\sim$2\% workflow-integration rate the published evidence reports, and is a constructed upper-contrast reference rather than a field-typical population. Real-world classification will face partially observed windows, rosters that churn mid-observation, and event streams of inconsistent quality across export formats --- none of which the current evaluation exercises.

\subsection{Schema and measurement limits.}\label{schema-and-measurement-limits.}

The event schema is deliberately minimal, and its proxies have known failure modes. With no task-type field, ``multi-step sessions'' proxies workflow depth by turn count --- but long sessions are ambiguous: a user deeply integrating a tool and a user struggling with it both produce many turns. The success-rate gate separates these where outcome labels exist, but task outcomes are deployment-reported and of heterogeneous quality; where success labels are sparse or unreliable, the Transform classification inherits that unreliability. The usage-regression flag measures what the schema can see --- declining invocation frequency --- and not what the construct wants: regression from achieved depth. The current simulator cannot even generate the latter pattern (its decay pathology suppresses frequency, never depth), so the framework can neither detect depth regression nor test for it; we name the flag for what it measures and record the sharper construct as future work. Finally, Notice-stage detection is only as good as the provisioning roster: incomplete rosters, contractors, and mid-window joiners and leavers all blur the framework's one count of the population that leaves no behavioral trace. The schema also records that an invocation occurred and whether it succeeded, but not what became of the agent's work --- whether a human accepted it, overrode it, or was escalated to; for human-in-the-loop agents that disposition is the discriminating signal, and we take it up in Section 8.5.

\subsection{Scope boundaries.}\label{scope-boundaries.}

NANTE as specified classifies one cohort's relationship to one deployed capability. Cross-tool aggregation --- a population adopting several AI systems at once, substitution among them, portfolio-level views --- is unresolved. The minimum-size and minimum-window floors mean small teams and young rollouts are unmeasurable by design; this is the honest price of cohort-level statistics, and it excludes exactly the small-team settings where qualitative observation works best --- a complementarity we consider a feature, but a boundary nonetheless. And the proposed thresholds are almost certainly not universal: what constitutes deep use plausibly varies by function, industry, and task type, which means single-organization findings will not transfer without the cross-organizational baseline described in Section 8.

\subsection{Effects of measurement.}\label{effects-of-measurement.}

Cohort-level aggregation addresses surveillance, but not Goodhart's law. A population that knows its adoption is measured can perform adoption: invocations without reliance, long sessions without integration. NANTE's depth-weighted constructs are harder to satisfy theatrically than activity counts --- sustained multi-week continuity with genuine task success is expensive to fake --- but not impossible, and any organization that converts NANTE scores into targets rather than diagnostics should expect the measure to degrade accordingly. Relatedly, the framework measures behavior, not experience: it can locate where a population stalled but not why in the participants' own terms. Perception instruments --- the surveys and assessments of Section 2.4 --- capture what telemetry cannot, and the framework is designed to sit beside them, not replace them.

%% file: sections-tex/07-incumbents.tex
\section{Why Existing Systems Have Not Bridged This}\label{why-existing-systems-have-not-bridged-this}

Section 2 established that the four measurement traditions rarely adopt one another's constructs; Section 2.5 argued the gap between them is structural. Here we make that argument explicit, because it bears on whether the gap should be expected to close on its own. Our claim is that each institutional position from which the unification could plausibly come --- the owners of the four traditions, and the integrator tier that serves them all --- is disposed against building it, for reasons that are stable rather than accidental.

\textbf{Evaluation and observability vendors} sell to engineering organizations, and their roadmap competes on the fidelity of system measurement: better traces, better evals, better cost attribution. Population-level behavior change is outside both their buyer's mandate and their own competence; the disciplines that would be required --- organizational psychology, change practice --- have no representation in their teams or their category's discourse. The recent appearance of end-user feedback and per-user views in these platforms extends the system's instrumentation to include humans as signal sources; it does not import a theory of organizational change, and there is no commercial pressure for it to.

\textbf{Platform vendors} face a subtler obstacle than is usually claimed. The convenient argument --- that adoption diagnosis threatens renewal --- is refuted by the vendors' own recent behavior: Microsoft now publishes habit-based user segmentation and benchmarks adoption across organizations, precisely because shallow deployment threatens expansion more than honest measurement does. What a platform vendor cannot credibly supply is \emph{neutrality}. Its instruments diagnose its own product, in constructs it defines, revisable at its discretion, scoped to its own ecosystem. An instrument that supports a substitution decision --- measure one population across this vendor's product and its alternatives, on identical open criteria, and act on the difference --- is one no platform vendor has an incentive to build, and none has built --- nor have the cross-vendor aggregators, whose common denominator remains the activity count. The adopting organization's interest is exactly that portfolio-neutral, auditable instrument; the vendor's interest ends at its own product's depth.

\textbf{The product-analytics tradition and its enterprise wing} --- growth analytics and customer success --- instrument employed populations with genuine sophistication, and are structurally committed to the wrong seat. Their practitioners are employed by vendors; their tooling is sold per account; their objective function is the renewal. The adopting organization's transformation office is not their market, and the organizational layer --- sponsorship, incentives, workflow authority --- is not in their models, which are pitched at the individual user and the funnel. Nothing prevents the instruments themselves from being repointed at an organization's own internal deployments, and occasionally they are; but repointing the tooling does not supply the change model, and the discipline that owns the tooling has no commercial reason to build one for a buyer it does not serve. The digital-adoption platforms, which do sit on the adopting organization's side, supply the seat without the model, per Section 2.3: instrumentation and guidance, with no computed staging of a population through a change model and no open thresholds to dispute.

\textbf{The change-management discipline} has the theory and the client relationship, but its economics are the economics of expertise: practitioner time, certification, and methodology licensing. Instruments that compute diagnoses from telemetry sit outside its delivery model and its skill base, and its established assessment products represent sunk intellectual property in the survey paradigm. A telemetry instrument is not an extension of that business; it is a partial substitute for it, offered in a medium the discipline does not produce in.

\textbf{Systems integrators and consultancies} occupy the last position from which the unification could plausibly come, and their economics argue against it differently. Adoption dashboards are built inside transformation engagements today --- bespoke, per client, billed as services --- and that is the form the delivery model rewards: instrumentation as deliverable, rebuilt each engagement, never productized into an open instrument whose thresholds a client could dispute or a competitor adopt, and never pooled into the cross-organizational baselines that would make any single engagement's readings comparable. The capability exists in this tier; the incentive to convert billable artifact into shared instrument does not.

\textbf{The professional separation compounds the commercial one.} The four traditions train different people, publish in different venues, reward different skills, and rarely employ anyone fluent across two of them, let alone four. The unification requires simultaneous working knowledge of agent evaluation practice, enterprise telemetry plumbing, behavioral analytics method, and staged change theory --- a combination that the current structure of these professions produces rarely and by accident.

We note the honest counter-trajectory: the boundaries are creeping. Workforce analytics now publishes behavioral staging; platform dashboards now export intensity metrics and benchmark across organizations; observability platforms now ingest end-user signals. The raw materials of adoption telemetry are increasingly present in every tradition. What the structure above predicts is not that the materials will stay separate, but that the \emph{interpretation layer} --- an explicit change model, computed from the adopting organization's own telemetry, in the adopting organization's interest, published openly enough to be audited --- is unlikely to come from any incumbent, because for each of them it is outside their competence, incompatible with the neutrality their position forecloses, or corrosive to their delivery model. Gaps of this shape are typically closed from outside the incumbent structure, which is what this paper attempts.

%% file: sections-tex/08-agenda.tex
\section{Research Agenda and Future Work}\label{research-agenda-and-future-work}

Each item below converts a limitation of Section 6 into a program of work. We begin with the items bearing most directly on the framework's central open question --- validity.

\subsection{Outcome validation.}\label{outcome-validation.}

The first requirement is evidence that NANTE's classifications mean what they claim. The study design is straightforward to state: instrument real deployments, classify cohorts prospectively, and test the classifications against independently observed outcomes --- whether cohorts classified as Transform or Embed in fact sustain changed work practices over subsequent quarters; whether cohorts diagnosed with a given stall type respond differentially to the intervention class mapped to it versus alternatives; whether the composite score predicts outcome measures the organization already trusts. Intervention-response validation is the strongest of these tests, because it examines the framework's practical claim --- that stall types are \emph{different kinds of problems} --- rather than only its descriptive one. Threshold calibration follows from the same data: fitting stage boundaries to observed outcome discontinuities rather than to proposed defaults.

\subsection{A cross-organizational baseline.}\label{a-cross-organizational-baseline.}

The stall-detection design of Section 4.3 hard-codes one empirical belief: that the Navigate-to-Transform boundary is a near-universal cliff, so that ordinary performance there should read as at-risk rather than failing. With telemetry from multiple organizations, that belief becomes a measurable distribution, and the instrument's verdicts can shift from absolute thresholds to position against the field: a cohort's depth conversion read as a percentile among comparable deployments rather than against a fixed floor. This is the framework's most consequential future direction --- it replaces its most contestable design decision with data, it makes verdicts self-updating as the field's own distribution moves, and it is only possible with pooled, anonymized, cohort-level curves across organizations. We note the state of the art precisely: Microsoft's Copilot Dashboard now benchmarks organizations against comparable companies, but by its own documentation ``benchmarks are only available for the percentage of active Copilot users, and for at most 6 months of historical data'' \citep{microsoft-adoption-report} --- cross-organizational comparison exists for breadth alone. The open problem is benchmarking \emph{stage progression}, which requires the shared change model those dashboards lack.

\subsection{Schema and simulator extensions.}\label{schema-and-simulator-extensions.}

Three concrete gaps from Section 6.3 define the near-term engineering agenda: a task-type field, so workflow breadth can be measured directly rather than proxied by turn counts; a depth-regression construct --- and the simulator pathology needed to test it --- so the framework can distinguish a population going quiet from a population losing achieved integration; and mixed-pathology simulation with threshold sensitivity analysis, so the evaluation of Section 5 can characterize the instrument's behavior under the entangled conditions real cohorts present rather than only under clean single-pathology populations. To these we add a fourth item from Section 5.1: the readiness-layer gate harness --- currently specification without implementation --- completing the framework's third layer in code.

\subsection{Ingestion of production export formats.}\label{ingestion-of-production-export-formats.}

The framework's substrate argument --- that the behavioral signal already exists in enterprise systems --- obligates adapters that consume those systems' actual exports: platform metrics exports at their published granularities, agent-platform analytics, and observability traces. This is engineering rather than research, but it is the engineering on which every study in 8.1 depends. These adapters must also confront an uneven field: production sources differ in which stages they can populate. Agent platforms log sessions and dispositions natively --- turn counts and task outcomes, the full schema, and with it every stage. Seat-assistant admin exports carry breadth and continuity --- sufficient for Notice through Navigate, and for Embed's continuity fraction --- but no task outcomes, leaving Transform's gates uncomputable from those exports alone. The instrument therefore degrades gracefully by design: it computes the stages a source supports and reports the remainder as unavailable, per the same convention as the insufficient-evidence rule of Section 4.6. We note the regularity this creates: Transform-grade telemetry exists precisely where organizations have built genuine agents rather than distributed seats, so the availability of depth measurement is itself a coarse indicator of deployment maturity.

\subsection{The disposition dimension.}

Unlike the items above, this subsection specifies a construct rather than proposing an experiment; we mark it as specification, not near-term agenda. The event schema of Section 4.2 records that an invocation occurred and whether its task succeeded. For human-in-the-loop agents, that is the wrong resolution: what distinguishes deep adoption from shallow is not whether the population invoked the agent but what happened to the agent's work --- whether it completed autonomously without review, was reviewed and accepted, was overridden by a human who preferred their own answer, was escalated to a human by design, or was abandoned mid-task. We propose extending the schema with a \emph{disposition} on each invocation across those five states, and letting disposition rather than turn count carry the depth construct. Which disposition signals depth is itself design-dependent, and the distinction matters: in a system built for human review, \emph{reviewed and accepted} is the healthy terminal state and unreviewed autonomous completion may indicate review bypass rather than success; in a system designed for autonomy, the ordering reverses. An instrument that cannot tell ratification from bypass cannot tell calibrated trust from rubber-stamping.

The construct is not ours. The organizational-behavior literature has converged on \emph{override rate} --- the proportion of AI recommendations a team reverses --- and \emph{override accuracy} as core indicators of calibrated trust \citep{atanassova2026override}, resting on a human-factors tradition that has distinguished appropriate reliance from misuse and disuse for three decades \citep{parasuraman1997,leesee2004}, and a measurement literature still working toward consensus on how reliance should be operationalized \citep{eckhardt2024}. Disposition telemetry is that construct computed from production signals rather than elicited from participants --- the same move NANTE makes for stage progression.

Four failure signatures follow directly, none detectable under the current schema. A cohort that invokes the agent, receives good output, and systematically overrides it exhibits \emph{trust failure without abandonment} --- invocations present, sessions deep, outcomes nominally successful, and no work actually delegated; under the present instrument this cohort can read as healthy. A cohort whose every non-trivial task escalates to a human exhibits \emph{guardrail-driven pseudo-adoption}, which the present instrument would likely misread as a shallow plateau and treat with workflow redesign when the constraint is permission scoping. And a cohort with high autonomous completion and declining invocation exhibits \emph{successful delegation}, which the present instrument reads as decay. A fourth cohort, whose work completes autonomously at a high rate with review rates falling over time, exhibits \emph{rubber-stamping} --- the over-reliance failure, in which agent output ships unexamined and defects reach production unremarked. This signature deserves particular emphasis because it is the one the present instrument would score most favorably: high completion, few escalations, no overrides. The reliance literature we draw on treats over-reliance and under-reliance as co-equal pathologies \citep{parasuraman1997,leesee2004}; an instrument that measures only whether people use a system, and not whether they examine what it produces, is equipped to see one of them. The first two are misdiagnoses, the third is an inversion, and the fourth is invisible in the direction that flatters the deployment.

Two things stand between this specification and an implementation, and we state both. The first is operational: the five states are defined here at the level of construct, not of computable predicate. Section 4.2's stages are predicates over named fields; \emph{override} is not yet that --- recognizing that a human preferred their own answer requires joining a later human action to an earlier agent output, a join that production systems frequently do not record and that would have to be specified per deployment before any threshold could be proposed. The second is evidentiary: we could not evaluate these signatures as we evaluated the five pathologies of Section 5. Those profiles were parameterized against published aggregate statistics on enterprise AI usage; no equivalent population statistic exists for agentic disposition. The adjacent literature offers reference points but not a base rate: clinical decision-support override rates pool near 90\% across studies with a 49--96\% range and extreme heterogeneity \citep{felisberto2024}, but they measure dismissal of interruptive alerts rather than reversal of delegated work; controlled studies of clinician reliance on labeled advice report susceptibility to incorrect recommendations between 27\% and 42\% depending on expertise \citep{gaube2021}; and vendor-reported autonomous-resolution rates for deployed customer-service agents span roughly 40\% to 85\%, with escalation rates from 2\% to 26\% --- figures that are marketing claims with inconsistent definitions of resolution rather than independent measurement. Simulating a disposition pathology from these would mean inventing the parameters and then verifying that our instrument recovers our invention --- precisely the circularity Section 5.4 declines. The disposition dimension is therefore specified here and left unimplemented on both counts --- not operationalized to predicate level, and not calibratable from anything published --- and it is the clearest case for the design-partner program that follows: the deployments that record dispositions at all, and the data that would calibrate them, exist only inside organizations currently running agentic systems, and nowhere else.

One consequence bears on the scope boundary of Section 3. That boundary is drawn on human initiation, because human-initiated invocation is what the present schema counts --- which places autonomously triggered systems outside it. Disposition is drawn on a different axis: what became of the work. A scheduled or triggered workflow emits dispositions even when no human initiated it, and the reference deployments whose escalation behavior we cite above are largely of that kind. We therefore expect disposition telemetry to be the construct through which some of the autonomous tier becomes measurable again --- not as a population progressing through a change, which it is not, but as a process whose delegation can be observed and whose failures of trust can be located. Whether that recovers enough of the third tier to constitute an instrument, or merely enough to constitute a research question, is exactly what production data would settle.

\subsection{Pairing behavioral and perception instruments.}\label{pairing-behavioral-and-perception-instruments.}

Section 6.5 positions surveys and practitioner observation as complements; the research question is sharper than coexistence. Practitioner accounts suggest that verbal and affective signals --- confidence hedges, permission-seeking --- precede behavioral divergence within teams. Whether perception instruments systematically lead telemetry as stall indicators, at what interval, and whether the two disagree in characteristic ways, is an open empirical question with direct operational value: it would define which instrument to trust at which altitude of the organization.

\subsection{An invitation.}\label{an-invitation.}

Every item above except 8.3 requires what no synthetic population can supply: production telemetry with observed outcomes. We invite organizations running enterprise AI deployments to participate as design partners, and participation is designed to survive contact with a privacy office. The instrument deploys in the partner's own environment --- the reference implementation is a local pipeline over the organization's own exports, and raw events never leave the organization's control. What a partnership contributes to the shared baseline is aggregate, cohort-level stage-distribution curves and outcome observations, anonymized at source under the constraints of Section 3; pooled curves will be held under a neutral governance structure defined with the founding partners, with minimum-cell rules applied to the pooled dataset as within any single organization; what a partner receives is its own cohorts' classifications, early access to the calibrated baselines the pooled curves make possible, and influence over the framework's development --- including the outcome-instrumentation design on which the studies of 8.1 depend. For partners running agentic deployments specifically, the collaboration extends to the disposition schema of Section 8.5: its five states are specified but uncalibrated, and only production disposition data can establish what their distributions look like in a healthy deployment. Systems integrators and managed-service providers who run adoption programs across many clients are a natural partner class: the per-engagement dashboards they already build are, in aggregate, the cross-organizational dataset this agenda needs. Participation through an integrator flows through that integrator's client consents, and partnership terms are structured accordingly. Partners control what is published about their deployments. The instrument, its thresholds, and its evaluation harness are open source, and the validation program's results will be published in the same spirit.

%% file: sections-tex/09-conclusion.tex
\section{Conclusion}\label{conclusion}

The failure of enterprise AI adoption is one of the best-documented phenomena in contemporary technology --- and one of the least instrumented. The literature that established the failure also established its cause: organizations do not lack capable systems; they lack changed behavior. Yet the measurement available to the organizations attempting that change describes everything except the change itself. Evaluation platforms report whether the system works. Usage dashboards report whether people touched it. Surveys report how people feel about it. No available instrument reports where, in the process of changing how work is done, a population has stopped --- as an auditable computation from the organization's own telemetry, through an explicit model of change.

This paper proposed that such an instrument is buildable from data organizations already possess. We defined \textbf{adoption telemetry} --- the continuous measurement of behavior change from production system signals, interpreted through an explicit model of change --- and specified the properties that distinguish it from its four neighboring traditions: computable, model-explicit, population-level, and intervention-mapped. We presented \textbf{NANTE}, a five-stage operationalization in which each milestone of a staged change model is expressed as a defined predicate over an event stream, with a stall-detection design calibrated to the field's actual distribution and a mapping from each diagnosable stall to a class of response. And we released an open-source reference implementation that demonstrates the framework's central technical claim: given only events and a roster, the instrument separates a healthy cohort from five characteristic failure modes --- including the stalled cohort that no activity metric can tell apart from it.

We have been deliberate about the boundary of the evidence. What is demonstrated is computability with discrimination on synthetic populations; what remains undemonstrated is validity on real ones. The thresholds are proposed, the pathologies are constructed, and the calibration that would convert structured hypotheses into findings requires production deployments with observed outcomes --- a program we have specified and for which we have invited partners. We consider stating this boundary plainly to be a feature of the work rather than a qualification of it: a measurement framework whose authors overstate its evidentiary status is a poor foundation for a discipline whose subject is the gap between appearance and reality in organizational behavior.

The structural analysis of Section 7 suggests this instrument was unlikely to come from the incumbents of any adjacent tradition, and the trajectory of the field suggests the need for it will grow rather than shrink: as agents become easier to build and easier to use, the binding constraint on their value migrates further from the technology and deeper into the organization --- from whether the system works, to whether work changes. The organizations that navigate that shift will be the ones that can see it happening. Our contribution is a first instrument for seeing --- offered openly, with its assumptions exposed, so that it can be disputed, calibrated, and improved by the communities whose separated expertise it attempts to unify.